\documentclass[twocolumn,american,aps,revmod,superscriptaddress,nofootinbib,showpacs]{revtex4-1}
\usepackage[english]{babel}
\usepackage[T1]{fontenc}
\usepackage{setspace}
\usepackage{outlines}
\usepackage[toc,page]{appendix}
\usepackage[colorinlistoftodos]{todonotes}
\usepackage[colorlinks=true, allcolors=blue]{hyperref}
\usepackage{amsmath}
\usepackage{xcolor}
\usepackage{placeins}
\usepackage{booktabs}
\usepackage{float}

\makeatletter
\renewcommand{\l@section}{\@dottedtocline{1}{1.em}{1.9em}}
\renewcommand{\l@subsection}{\@dottedtocline{2}{2.0em}{1.5em}}
\renewcommand{\l@subsubsection}{\@dottedtocline{3}{3.em}{1.5em}}
\makeatother

\begin{document}
%\begin{frontmatter}

%\title{Physics Based Loss Function on the Surrogate of NLTE Model for ICF Simulations}
%\title{Evolution of Ionization in Laser-induced Steady-State Plasmas}

\title{Ionization Dynamics in Intense Laser-Produced Plasmas}% Force line breaks with \\

\author{M. S. Cho}
\email{cho28@llnl.gov}
\affiliation{Lawrence Livermore National Laboratory, 7000 East Avenue, Livermore, California 94550, USA}
\author{A. L. Milder}
\affiliation{Laboratory for Laser Energetics, 250 East River Road, Rochester, New York 14623, USA}
\affiliation{Department of Physics, University of Alberta, Edmonton, Alberta T6G 2E1, Canada}
\author{W. Rozmus}
\affiliation{Department of Physics, University of Alberta, Edmonton, Alberta T6G 2E1, Canada}
\author{H. P. Le}
\author{H. A. Scott}
\affiliation{Lawrence Livermore National Laboratory, 7000 East Avenue, Livermore, California 94550, USA}
\author{D. T. Bishel}
\affiliation{Laboratory for Laser Energetics, 250 East River Road, Rochester, New York 14623, USA}
\affiliation{Department of Physics and Astronomy, University of Rochester, Rochester, New York 14627, USA}
\author{D. Turnbull}
\affiliation{Laboratory for Laser Energetics, 250 East River Road, Rochester, New York 14623, USA}
\author{S. B. Libby}
\affiliation{Lawrence Livermore National Laboratory, 7000 East Avenue, Livermore, California 94550, USA}
\author{M. E. Foord}
\affiliation{Lawrence Livermore National Laboratory, 7000 East Avenue, Livermore, California 94550, USA}

\date{\today}
\begin{abstract}
The ionization dynamic of argon plasma irradiated by an intense laser is investigated to understand transient physics in dynamic systems. This study demonstrates that significant delayed ionization responses and stepwise ionization processes are crucial factors in determining the ionization state of such systems. When an intense laser begins to ionize an initially cold argon plasma, the conditions change rapidly, leading to a delayed response in ionization. Consequently, the dynamics do not reach a steady state, even if the electron temperature and density appear unchanged, particularly when the atomic transition process is not sufficiently rapid compared to the relevant time scales. Furthermore, in this case, numerous highly excited states are created primarily through collisional excitation. Thus, even low-energy photons can predominantly ionize plasmas, challenging the conventional belief that such photon energies insufficient to overcome the binding energy of bound electrons typically contribute less to the ionization. These findings underscore the necessity of incorporating these processes in ionization modeling within radiation hydrodynamic simulations for various laser-plasma experiments.
\end{abstract}

%\keywords{Suggested keywords}%Use showkeys class option if keyword
                              %display desired
\maketitle

% (1) Background
\sloppy % Allows more flexible line breaks

Ionization processes are crucial in determining a plasma's properties. These properties include, but are not limited to, electrical conductivity \cite{PhysRevE.58.6557}, heat transport characteristics \cite{farmer2020validation,PhysRevLett.124.025001,le2019influence}, and optical emission spectra \cite{de2017laser}. The ionization state also affects the sound speed and collisionality of the plasma. This impacts not only laser-plasma instabilities but also diagnostic techniques such as Thomson scattering \cite{sheffield2010plasma}. In these techniques it is often difficult to distinguish between changes in ionization state and temperature. The study of ionization dynamics thus becomes indispensable for comprehending the behavior of plasmas under laser irradiation. This knowledge is vital for applications ranging from material processing to fusion research, where control and predictability of plasma properties are essential \cite{falcone2024materials,hu2024review,hurricane2024energy}.

In this Letter, we elucidate two surprising phenomena in the ionization dynamics of intense laser-induced plasmas: a long delayed ionization response from rapidly changing conditions, and the dominance of a two-step ionization mechanism that involves collisional excitation followed by photoionization. Initially, as the plasma undergoes rapid changes in its conditions, the ionization fails to achieve a steady state at each timestep, leading to a lag in ionization. Notably, this effect can persist even when plasma conditions appear to stabilize, resulting in an ionization level that is consistently more than 15\% lower than expected under steady-state conditions. Additionally, despite photon energies (\(\sim 3.5\) eV) being below the typical electron binding energy threshold (\(\sim\) a few tens of eV), our simulations show that intense laser pulses primarily produce a photoionized plasma environment. This is facilitated by electron collisions that populate higher energy levels, leading to subsequent photoionization. In radiation hydrodynamics simulations for laser-plasma experiments, several methodologies for calculating ionization, which are strategically chosen for rapid computation, have yet to consider the importance of these physical phenomena, thereby highlighting their significance \cite{scott2022using, FRANK2022100998, kluth2020deep}.

To investigate the dynamics of the ionization process when a gas target is subjected to laser irradiation, Non-Local Thermodynamic Equilibrium (NLTE) calculations \cite{scott2010advances, chung2013comparison,mashonkina2005nlte,PhysRevE.109.045207} are performed, employing the experimental parameters from the Omega Laser Facility \cite{milder2022direct}. In these experiments, eleven laser beams, each imparting 200 joules of energy, are concentrated onto argon gas from a supersonic jet to investigate the heat transport mechanisms within the ensuing plasma. The flat-top laser pulse has a FWHM duration of 1 ns with rise and fall times near 100 picoseconds, and a peak intensity of roughly 1.1 x $10^{15}$ $W/cm^2$ at the focal point. A wavelength of 351 $nm$ is used, and the ion density in the plasma is estimated at $1.8$ x $10^{18} cm^{-3}$. Thomson scattering is utilized to measure electron temperature and density at the plasma's core, providing benchmarks for our simulations. 

Ionization calculations are conducted using the code Cretin \cite{scott2001cretin, scott2010advances}, based on the experimentally measured temperatures and densities. Cretin, extensively utilized in laser plasma experiments, especially for indirect inertial confinement fusion (ICF) simulations \cite{lindl2014review,hurricane2023physics}, proves to be well-suited for the analysis of this experiment. This code facilitates time-dependent modes capturing ionization dynamics, which allows comparison with the commonly used steady-state models in plasmas with $<$10\% critical density \cite{milder2021measurements, turnbull2023}. Note that the time-dependent mode operates in two distinct forms. The first involves calculating ionization based on inputs that reflect temporal changes in plasma conditions (TD-P), such as electron density ($N_e$) and temperature ($T_e$). The second method computes ionization by taking the characteristics of the laser used and the density of the target as initial conditions (TD-L). In this mode, $N_e$ and $T_e$ are calculated self-consistently within the framework of the code. (See 'Supplementary Material 1' for more detailed information on NLTE calculation modes.) 
%Detailed descriptions of these computational methodologies are provided in Appendix A.

%\begin{figure}
%    \centering
%    \includegraphics[width=.95\linewidth]{csd_v2.png}
%    \caption{\label{csd} Charge state distribution of the argon plasma as a function of time. [inset] Laser intensity as a function of time. The laser pulse is a 1ns flat-top  time duration (FWHM) with rise and fall time of 100 ps.}
%\end{figure}

\begin{figure}
    \centering
    \includegraphics[width=.95\linewidth]{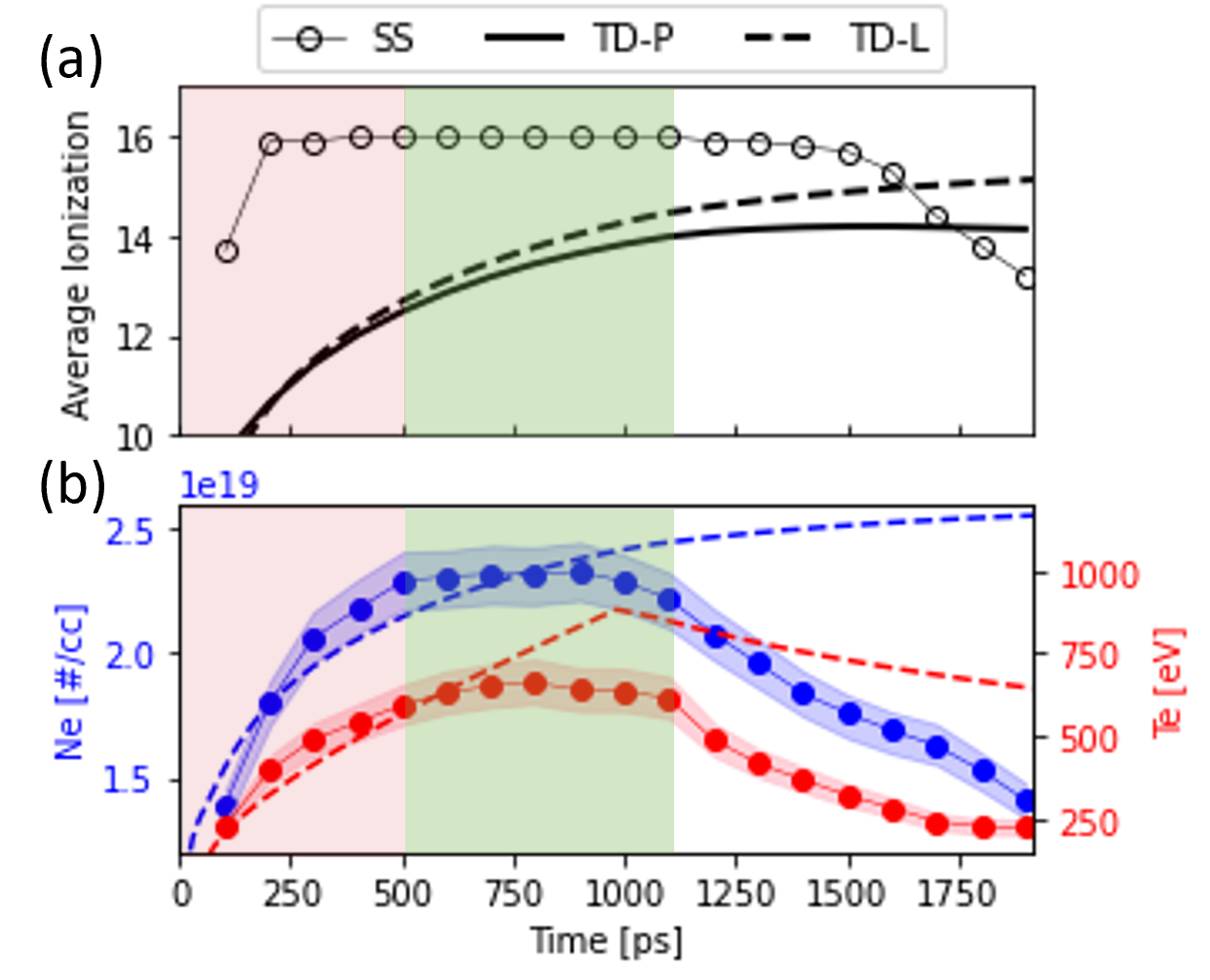}
    \caption{\label{ionization_curve} (a) The evolution of average ionization under various calculation modes is examined. Ionization assuming the steady-state (SS) condition is denoted by circles, in contrast to the time-dependent mode with plasma conditions (TD-P), which is represented by a solid line. The dashed line represents another time-dependent mode involving laser inputs (TD-L), which provides simulated $N_e$ as blue and $T_e$ as red dashed lines in (b). Measured $N_e$ and $T_e$ at the center of an Ar gas are plotted together as circles with experimental error bands, redrawn based on Ref.\cite{milder2022direct}. The rapidly changing plasma conditions are marked by a pink shaded region, and the green shaded region illustrates the quasi-stable state where $T_e$ and $N_e$ exhibit minimal variation due to the balance of incoming and outgoing energy in the system.}
\end{figure}

Figure \ref{ionization_curve}(a) illustrates the temporal evolution of ionization for Ar under measured plasma conditions, demonstrating significant lag in ionization response compared to the steady-state values within the NLTE model, despite identical plasma input conditions. In the steady-state assumption, the ionization degree is calculated by assuming plasma equilibrium at each measured point of temperature and density. Notably, at the initial measurement interval of 200 ps, the plasma displays an ionization degree of approximately 16, suggesting that all electrons, except those in the K-shell, are ionized. Conversely, the TD-P calculations, using identical plasma conditions, shows a persistence of many electrons in the levels with the principal quantum number (n) $>$ 1, failing to reach an ionization degree of 16 within the observed timeframe. This phenomenon indicates that in systems undergoing rapid plasma condition changes ($<$ 500 ps), the rate of ionization cannot match the pace of these changes, preventing equilibrium attainment, corroborating a model previously proposed by D. Bishel, et al. \cite{bishel2023ionization}. 

More importantly, the differences in ionization is still observed within the quasi-stable period, the green-highlighted region in Fig.\ref{ionization_curve}(a) where key plasma parameters, $N_e$ and $T_e$ remain stable. This plasma appears to maintain steady conditions, suggesting that the system and other dependent physical quantities might also remain stable. Nonetheless, time-dependent NLTE calculations indicate ongoing ionization processes within this segment. At equilibrium, the plasma conditions—specifically $N_e$ of about 2.3x$10^{19}$ cm$^{-3}$ and $T_e$ of 600 eV—are sufficient to ionize all electrons except K-shell electrons. However, given the rapid attainment of these conditions, the ionization rate fails to stabilize even by 1 nanosecond. In this period, the collisional ionization serves as the predominant ionization mechanism. When electron densities stabilize at approximately \(2.4 \times 10^{19}\) cm$^{-3}$, the collisional ionization rate afforded by the Lotz model implemented within Cretin \cite{lotz1968electron}, is $\sim$ \(10^{10}\) Hz. This rate is not sufficiently rapid to compensate for previously established disparities in ionization within this regime. Therefore, the free electrons in the system, during this temporal range, continue to ionize through collisions, driving the system toward a steady-state ionization degree under quasi-stable conditions.

For validation, we conduct a time-dependent calculation using the laser profile from the experiment and maintaining a constant initial ion density (TD-L). Calculations with a laser intensity of ($3$x$10^{14}$ $W/cm^{2}$) and initial gas density($1.8$x$10^{18}$ $cm^{-3}$) show agreement with the measured plasma conditions [Fig. \ref{ionization_curve}(b)] and TD-P ionization states [Fig. \ref{ionization_curve}(b)]. Like the TD-P calculations, the TD-L calculations show an evolution of the ionization state while the plasma conditions are quasi-stable. Conclusively, if the system's plasma conditions evolve rapidly, the ionization degree will continue to evolve both during and after this rapid transition phase. The ionization in the later phase (post-500 ps) is slightly lower than in TD-P calculation, due to the model's zero-dimensional nature, which does not account for spatial dynamics such as heat conduction or plasma expansion. This leads to slightly different \( N_e \) and \( T_e \) values that are calculated self-consistently in this phase.

Unlike simulation modes relying on $N_{e}$ and $T_{e}$ which obscure the direct influence of lasers on the ionization process, a notable advantage of the TD-L calculation is its ability to investigate the ionization processes induced by intense laser fields at the atomic level. Therefore, we analyze the ionization dynamics during the initial 500 picoseconds which provides good agreement of $N_{e}$ and $T_{e}$ with experiment, as shown in Fig.\ref{ionization_curve}(b), when the effects of heat conduction and hydrodynamic expansion are negligible.

\begin{figure}
    \centering
    \includegraphics[width=.95\linewidth]{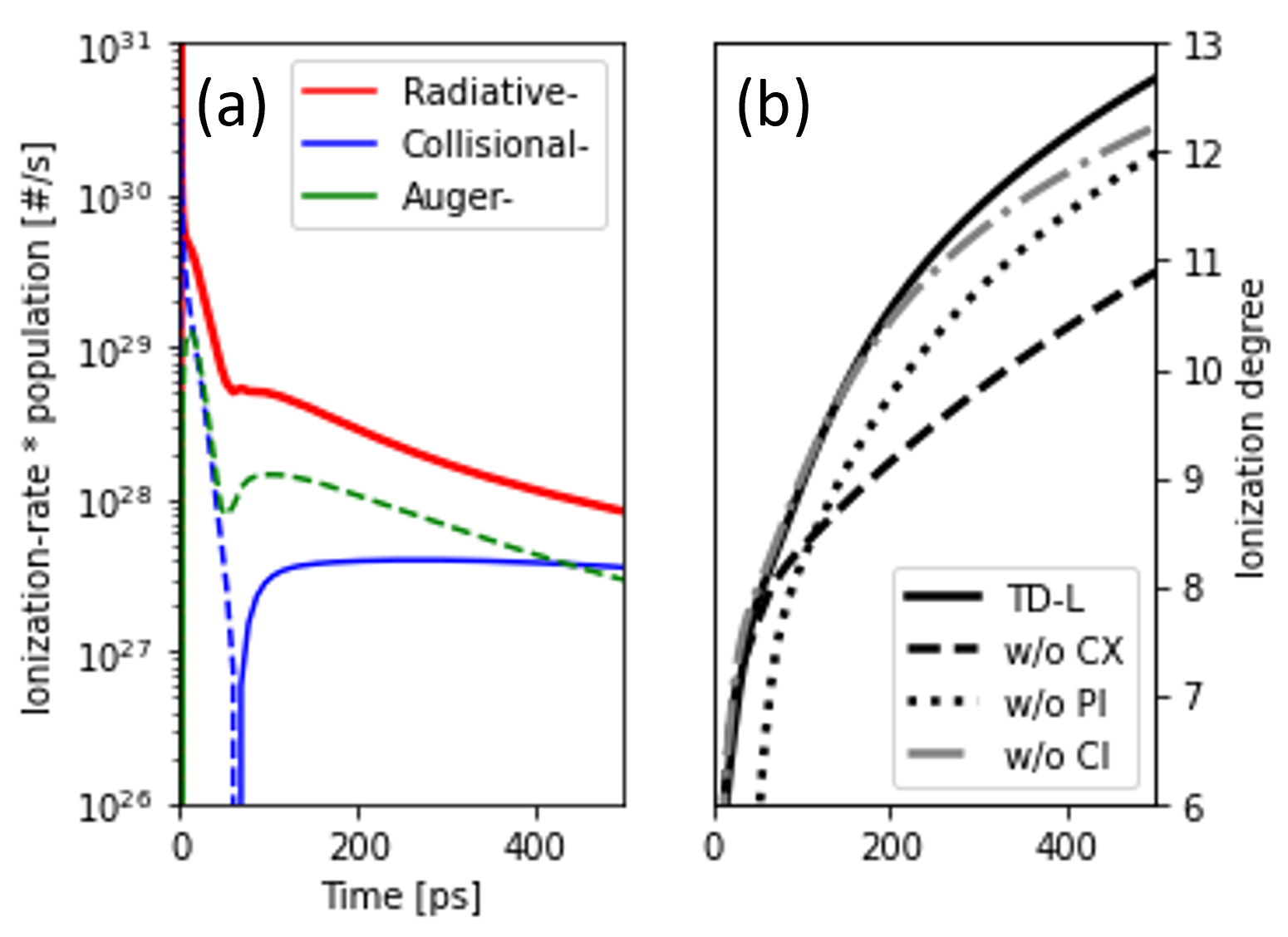}
    \caption{\label{ionization} (a) Net population flux of three different ionization processes in time: Radiative ionization process (red), collisional ionization process (blue), auto ionization process (green) considering their inverse processes. "Net" denotes the subtraction of the recombination processes (the reverse of ionization), with the absolute value taken, the dash line indicates the inverse process is dominant. (b) Detailed time-dependent ionization calculations for the first 500 ps, while omitting individual transition processes. This approach enables focused analysis on how each transition affects the overall ionization. 'w/o PI', 'w/o CX', and 'w/o CI' represents ionization of TD calculation without photo-aided ionization, collisional excitation, and collisional ionization, respectively. The inverse of each process is also turned off simultaneously for each case.}
\end{figure}

\begin{figure}
    \centering
    \includegraphics[width=.95\linewidth]{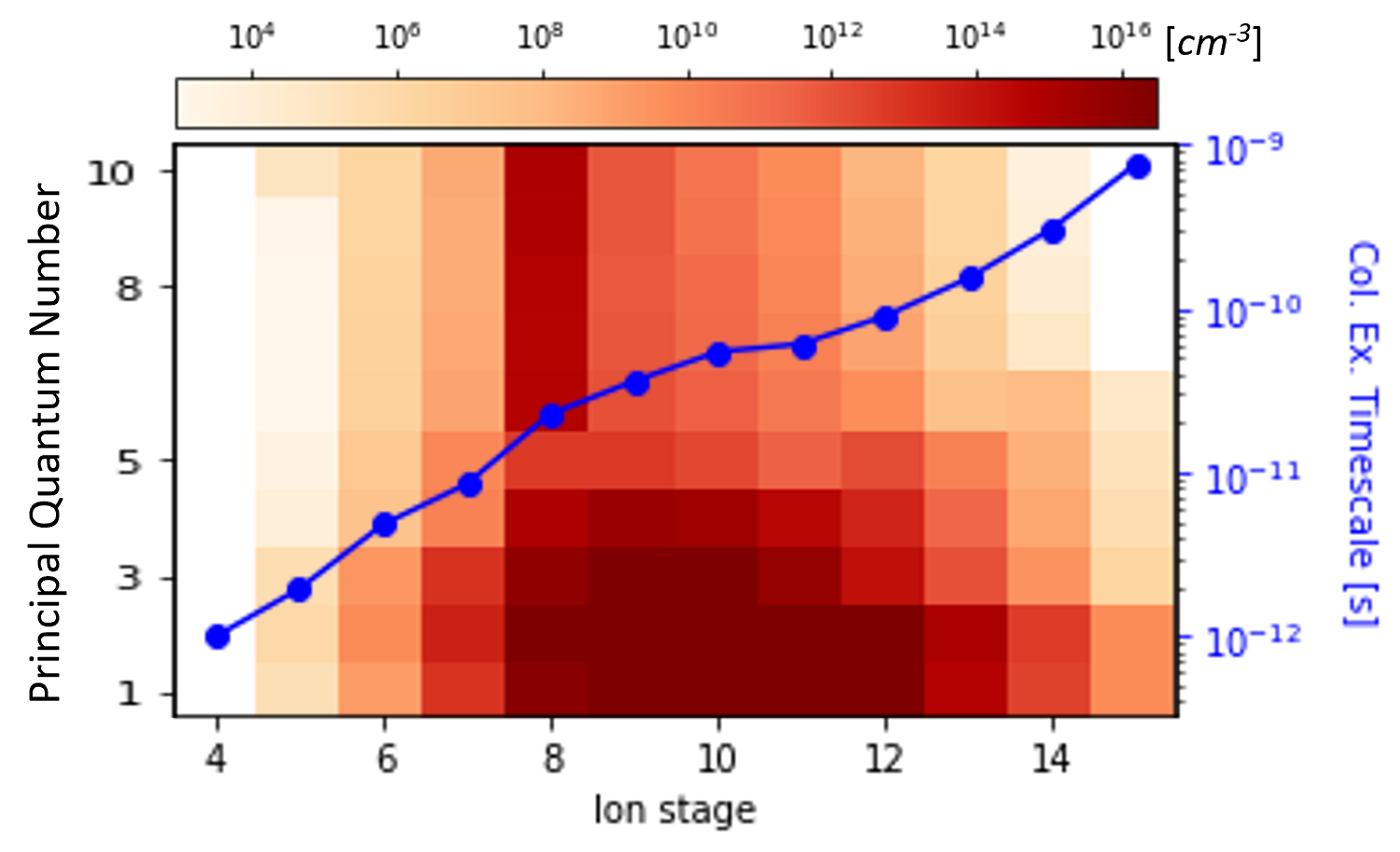}
    \caption{\label{pop_rate} Population histogram of bound electrons in relation to their ionization states and principal quantum numbers at 100 picoseconds. The collisional excitation timescale, which is the inverse of the collisional excitation rate, depending on the ion stage are depicted on the right axis. The timescale specifically accounts for the transition from the ground state to the first excited state of each ion, as this transition typically has the smallest rate among all excitation processes due to the largest energy difference.}
\end{figure}

The primary finding of our study is that during the initial 500 picoseconds, the two-step ionization mechanism, where excitation occurs first and is followed sequentially by ionization, plays a significant role in ionization. Figure \ref{ionization}(a) illustrates the net population flux, calculated by multiplying each ionization rate by the population of the corresponding configuration, thereby quantifying the transitions attributable to each ionization process over time. The simulation indicates that the radiative ionization process, or photo-aided ionization process—meaning ionization driven by photons, including multi-photon ionization—dominates. This finding is particularly surprising given that the photon energy in our system, approximately 3.5 eV, is substantially lower than the ionization potential of Ar plasmas (e.g., the ionization energy of $\text{Ar}^{6+}$ is greater than 100 eV). The bound-electron distribution, depicted in Figure \ref{pop_rate}, provides insight into this discrepancy. It shows the distribution of bound electrons across principal quantum numbers ($n$) for each ion at 100 picoseconds, indicating a significant population of electrons in high-$n$ levels accessible to photo-aided ionization. This suggests that excitation processes are simultaneously occurring in the system, populating high-$n$ energy levels with sufficiently low binding energies to be accessible via a few photons. For the reference, tunneling ionization rates are not included because the Keldysh parameter is approximately 1.5 under the system conditions, indicating that multiphoton ionization is the dominant process and thus primarily considered \cite{keldysh2024ionization}.

Furthermore, this excitation into high-n levels primarily arises from collisional processes. Figure \ref{ionization}(b) demonstrates this, as turning off the collisional excitation process significantly reduced ionization after 70 picoseconds. Compared to the minimal impact observed when collisional ionization process is disabled, turning off photo-aided ionization process reduces ionization, emphasizing it as the primary ionization mechanism. Turning off collision excitation results in the lowest ionization, as it prevents electrons from being excited to high n-levels, leading to a significant reduction in both photo-aided ionization and collisional ionization. 

Physically, the superior efficacy of photons over electrons for ionization of electrons in high-n levels can be understood by considering the number of free electrons and photons in the system. The cross-sections for both photoionization and collisional ionization at these levels, calculated respectively using Kramers' formula and Lotz's formula, are both $\sim$ $10^{-16}$ $cm^2$ and do not significantly differ \cite{kramers1923xciii, lotz1968electron}. Since the photon number flux, approximately $10^{33}$ $cm^{-2} s^{-1}$, vastly exceeds the electron flux in that range ($\sim$ $10^{28}$ $cm^{-2} s^{-1}$), photo-aided ionization is more likely to occur when the excited levels are populated. Additionally, despite the presence of a large number of photons, the collisional process becomes the primary excitation mechanism due to the high photo-deexcitation rate, which compensates for photo-excitation. 
%Note that the collisional excitation rate determines the ionization time scale since it has quite lower rate compared to the radiative ionization process in two step mechanisms. The collisional excitation rate, which varies for each transition, is found to be less than $10^{10}$ Hz for ground states of ions with a charge greater than +12. This lower rate ultimately leads to a delay in ionization in time-dependent calculations compared to steady-state calculations.

Additionally, within the two-step ionization mechanism, the collisional excitation rate governs the ionization time scale due to its substantially lower rate compared to the radiative ionization process. The time scale of the collisional excitation, which is the inverse of the transition rate, from the ground state to the first excited state of each ion is illustrated in Fig.\ref{pop_rate}. As ionization advances, the energy gap between these states widens, resulting in increasingly longer collisional excitation times. Consequently, for ions with a charge state of +7 or higher, the process can extend over tens of picoseconds, while for ions with a charge state of +12 or higher, it can extend over hundreds of picoseconds. Given that the rate of the second step, photo-aided ionization, generally occurs on a femtosecond timescale, this disparity is the primary factor contributing to the ionization lag.

% During this period, the Auger process primarily facilitates electron recombination (electron capture), while the collision process is dominated by three-body recombination during the first 100 picoseconds, subsequently transitioning to an increase in ionization. Each of the three curves exhibits an inflection point around 70 picoseconds, which corresponds to an average ionization state of approximately +8 at that time, marking the stage where all M-shell electrons are ionized. The L-shell's binding energy exceeds the M-shell's by over 100 eV, thereby decelerating electron capture. Three-body recombination, which requires free electrons to accelerate to the energy levels of bound electrons, becomes increasingly improbable.

Due to the complicated interplay that causes ionization lag, predicting the conditions of the ionization lag with simple physical parameters proves to be exceedingly challenging and requires time-dependent calculations. Therefore, we propose an empirical formula that mandates when time-dependent calculations are needed for various laser intensities and target densities - parameters readily measurable in experiments: 

\begin{equation} \label{empirical_formula}
    t_{eq} = \frac{5e^{-4I_{14}}+3}{N_{18}-ln{(\frac{4}{I_{14}}})} [ns]
    %t_{eq} = \frac{5*f(I_{14})+3}{N_{18}-ln{(\frac{4}{I_{14}}})} [ns]
    %t_{eq} = \frac{f(I_{14})}{N_{18}-ln{(\frac{4}{I_{14}}})} [ns]
\end{equation}

where $I_{14}$ is the laser intensity with the unit of $10^{14}$ $W/cm^2$ and $N_{18}$ is the ion density with the unit of $10^{18}$ $cm^{-3}$.
It elucidates the convergence duration of time-dependent NLTE calculations with steady-state calculations, informed by a set of 100 Cretin simulation cases. It is derived by first setting specific laser intensity and ion density density to perform TD-L calculations, then using the resulting $N_e$ and $T_e$ as inputs for SS calculations at each timestep. The convergence duration refers to the time when the ionization difference between these two calculation methods converges to within 5$\%$. For this formula, the laser is characterized by a flat top pulse shape with 1 nanosecond time duration, while the ion density remains constant over time, allowing only the electron temperature and density to evolve. (See 'Supplementary Material 2' for more detailed information on the derivation process.)

\begin{figure}
    \centering
    \includegraphics[width=.95\linewidth]{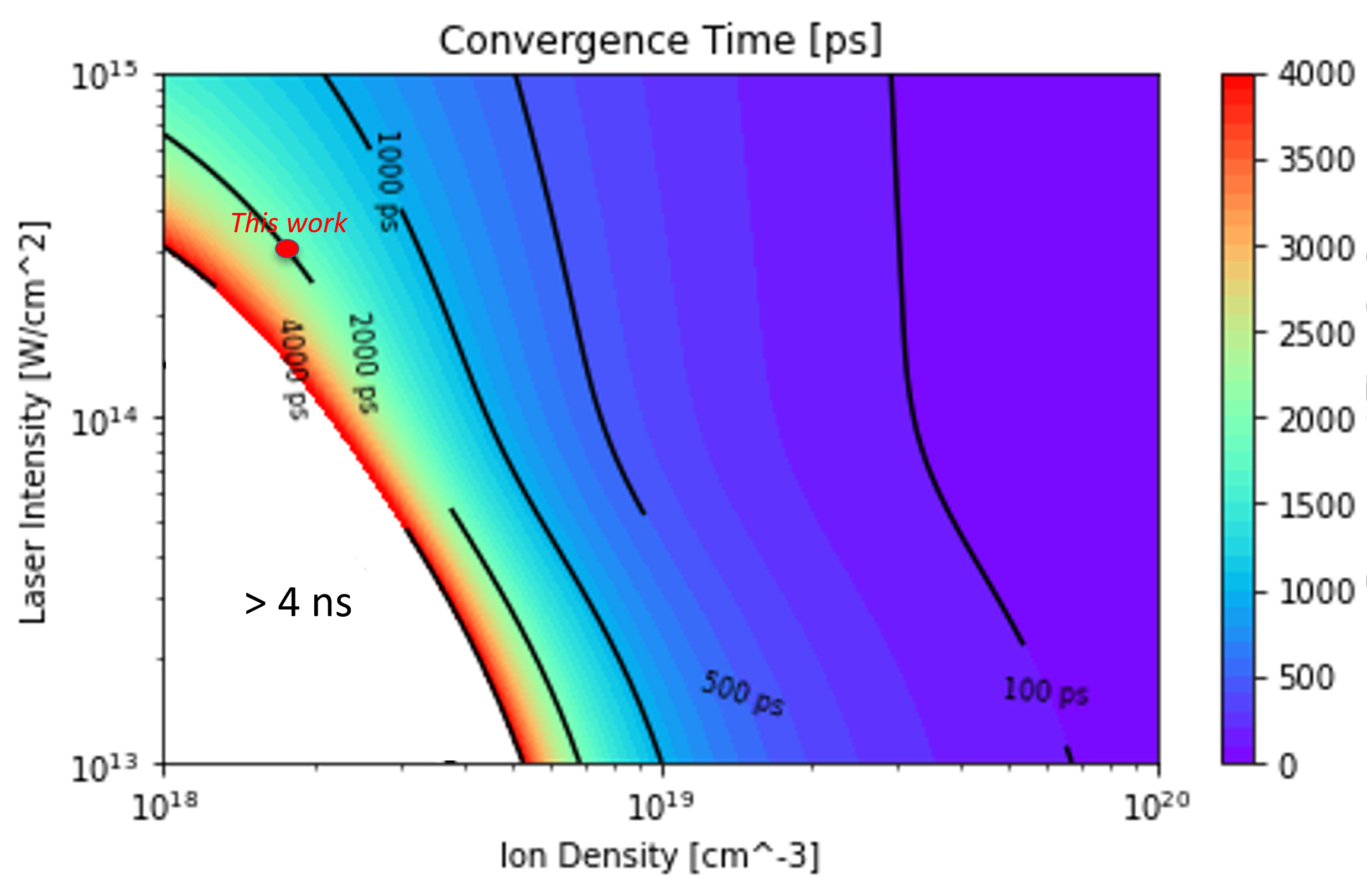}
    \caption{\label{formula_em} Convergence time as a function of laser intensity and initial ion density. Convergence time refers to the duration taken for the time-dependent calculations to converge to a steady state in terms of ionization. The white region marks the conditions where the convergence time is longer than 4 ns. This work is highlighted by the red dot.}
\end{figure}

This formula demonstrates that an increase in density results in a higher frequency of collisional processes, thereby reducing the collisional excitation timescale, which is inversely proportional to the density. Similarly, an increase in laser intensity raises the temperature and the number of free electrons, consequently reducing the collisional excitation timescale as well, which decays exponentially with the intensity in the numerator of the formula. In summary, the empirical formula suggests that for systems where the calculated convergence time exceeds the hydrodynamic timescale, ionization lag will occur. This necessitates the use of a time-dependent model to accurately predict ionization degrees. 

Figure \ref{formula_em} depicts the variation in convergence time derived from this formula. Our analysis shows that a convergence time of approximately 2 nanoseconds is necessary for the calculated values from both modes to align, in this study. This highlights that time-dependent modes inherently yield ionization dynamics different from those of the steady-state mode, requiring time-dependent calculations. Conversely, experiments with higher densities and laser intensities, which involve shorter timescales, can be sufficiently modeled using the steady-state mode. 

For the practical usage of this formula, there are a few caveats: this model is specific to Ar gas targets, and similar trends will be observed for other materials, albeit with different corresponding laser intensities and gas densities \cite{Turnbull2024pop}. Additionally, it does not account for heat conduction and radiation-driven hydrodynamic expansion, which influence dynamics on the nanosecond timescale and ultimately extend the convergence time. Moreover, densities from \(10^{18}\) to \(10^{20}\) \(\text{cm}^{-3}\) and laser intensities from \(10^{13}\) to \(10^{15}\) \(\text{W/cm}^2\) are used in the derivation, suggesting that the formula's applicability is confined to these ranges. Therefore, this formula should be used as an approximate guide for scenarios where time-dependent NLTE calculations are necessary, especially in comparison to the timescales of interest to experimentalists.

In conclusion, the ionization dynamics of Ar plasma under intense laser irradiation, as investigated by a NLTE model, reveal an ionization lag not only during rapid changes in plasma conditions but also under subsequent steady-state conditions. The transition process in these conditions is not faster than this timescale, which causes the lagging and deviations from steady-state predictions. Additionally, a significant finding is the dominance of photoionization in the ionization process, challenging the conventional belief that low photon energy is insufficient for ionization. This enhances our understanding of ionization mechanisms and informs the strategic application of NLTE models in plasma studies. The empirical formula derived from Cretin simulation analyses aids in selecting computational approaches based on ion density and laser intensity variations.

This work was performed under the auspices of the U.S. Department of Energy by Lawrence Livermore National Laboratory under Contract No. DE-AC52-07NA27344. This material is also based upon work supported the Department of Energy [National Nuclear Security Administration] University of Rochester “National Inertial Confinement Fusion Program” under Award Number(s) DE-NA0004144.

\bibliography{ref}

% Here begins the supplementary material
\clearpage
\twocolumngrid
\appendix
\section*{Supplementary Material}

\subsection{Three Modes of NLTE Calculation}
% Detailed experimental procedures that are supplementary

%\label{sec:appendixA}
%\section{Three modes of NLTE calculation: Steady-state(SS), Time-dependent mode with plasma conditions (TD-P) and with laser conditions (TD-L)}

\setcounter{figure}{0}
\renewcommand{\thefigure}{A-\arabic{figure}}

NLTE model provides population distributions and characterizes the physical processes occurring in the plasma for a given electron temperature and density. NLTE model solves sets of rate equations to calculate the number density of the $i^{th}$ atomic state $n_i$ as a function of rates $A_{j \rightarrow i}$ from the jth atomic state to the $i^{th}$ atomic state, where 1 $\le$ i, j $\le$ m (the maximum number of atomic states) from the following equation.

\begin{equation} \label{rate_eq}
    \frac{dN_{i}}{dt} = \sum_{j \neq i} ( A_{j \rightarrow i}N_{j} - A_{i \rightarrow j}N_{i} ) 
\end{equation}

where $N_{i}$ represents the state population at the state $i$ and $A_{i->j}$ means the transition rates from the state $i$ to $j$,  describing how quickly the transition occurs. They include plasma processes, e.g. photon-excitation and ionization, collisional-excitation and ionization, autoionization and their inverse processes obtained by the principle of detailed balance. Typically the rate equation manifests as a matrix equation of high dimensionality. To present this more concisely, it can be reformulated in a matrix representation, drawing upon Eq. \ref{rate_eq}:

\begin{equation}
\frac{d}{dt} 
\begin{bmatrix}
n_1 \\
\vdots \\
n_M 
\end{bmatrix}
= 
\begin{bmatrix}
A(n_1, T_e, n_e, J) \\
\vdots \\
A(n_M, T_e, n_e, J)
\end{bmatrix}
\begin{bmatrix}
n_1 \\
\vdots \\
n_M 
\end{bmatrix}
\label{eq:matrixform2}
\end{equation}

where the $n_x$ is the population of the $x_{th}$ energy level. The solution to the rate equation is derived by utilizing known parameters of the plasma state. Specifically, in situations where the plasma is in a steady state, it exhibits no temporal variations in population levels within the rate equation framework. As a result, the differential term that represents population change over time, located on the left side of the equation, is set to zero to simplify the solution process. Critical inputs at this stage include the plasma's electron temperature ($T_e$) and density ($n_e$), which are used to calculate the transition matrix. Additionally, the radiation field $J$ includes the external radiation field information $J_{\text{ext}}$, in this case, the laser information, and the self-generated radiation field $J_{\text{self}}$, which can be excluded under the assumption of optically thin plasmas in the calculation of the transition matrix $A$. Note that although the external radiation field $J_{\text{ext}}$ serves as a main heating source for the plasmas, when we set $T_e$ in the calculation, which already accounts for the outcomes of laser heating, including $J_{\text{ext}}$ would result in double-counting the incident energy. Therefore, the resulting population distribution reflects the ion distribution within the plasma, characterized by a steady state under specific temperature and density conditions only.

Conversely, in scenarios where plasma temperature and density exhibit temporal variations, the differential term of the population with respect to time becomes non-zero. Under these dynamic conditions, the transition matrix $A$ undergoes continuous adjustments in response to the evolving temperature and density, significantly influencing the resulting population distribution. The approach to computing electron temperature, $T_e$, necessitates the use of two distinct subroutines. If $T_e$ and electron density, $n_e$, are supplied as external, time-dependent inputs, the population is derived accordingly (TD-P). Thus, the population from the preceding time step serves as the initial condition for the subsequent step and is subsequently updated by multiplying with the transition matrix. Conversely, when initial values of $T_e$ and $n_e$ are provided along with other variable conditions, such as the temporal history of the input radiation field (e.g., laser temporal profile), the population change is initially determined by the radiative transition rate (TD-L). This, in turn, allows for the self-consistent calculation of $T_e$ and $n_e$ based on the principles of energy conservation. Technical details for NLTE population kinetics is described on the Ref. \cite{scott2001cretin, chung2005flychk,scott2010advances,ralchenko2016modern,cho2018intensity,cho2020opacity}.

%\begin{figure*}[!ht]
%    \centering
%    \includegraphics[width=.95\linewidth]{FigrB_2_1.png}
%    \caption{\label{trho_plot} Comparison of simulated (a) $T_{e}$ and (b) $N_{e}$ with the %measurement. Three cases shows the temporal evolution depending on different initial ion %density with the unit of $cm^{-3}$. The electron density looks very sensitive to the initial %ion density.}
%\end{figure*}

The TD-L mode of the NLTE simulation used in this work lacks spatial information, thus it cannot fully capture plasma dynamics based solely on laser intensity and initial ion density data. For instance, it does not account for theoretical predictions or factors such as laser absorption, heat and radiation transfer, or plasma expansion. To address this limitation, we assumed variations in actual laser intensity or initial ion density within an expected range and aimed for temperature and density calculations that closely matched experimental observations. This approach is grounded in the use of density and temperature as primary quantities in the rate matrix for ionization calculations in the NLTE model.

%Figure \ref{trho_plot} juxtaposes electron temperature and density derived from the TD-L mode against experimental data. Despite using a peak laser intensity of $1 \times 10^{15}$ W/cm$^2$ in experiments, which varies spatially, and employing laser absorption with a Langdon reduction factor based on the Lee-More model, ongoing research is refining this factor \cite{LeeMore,scott2010advances,turnbull2023}. Hence, laser and ion density parameters were systematically scanned around $1 \times 10^{15}$ W/cm$^2$ and $1.5 \times 10^{18}$ ion density to optimize density and temperature predictions. This exploration yielded optimal values of $3 \times 10^{14}$ W/cm$^2$ for laser intensity and $1.8 \times 10^{18}$ for density, aligning well with initial ion and temperature generation. Notably, the plasma's sound speed is approximately $\sim 1.6 \times 10^7$ cm/s, with the plasma expanding by over 20\% in volume within 500 picoseconds at this speed \cite{NRLformulary2019}. Accordingly, conditions were set to match temperature and density within the initial 500 picoseconds. Despite the uncertainty and limit of NLTE simulation mentioned above, it does not change the main result of this work; ionization simulations failed to reach steady-state values within 1 nanosecond. Figure \ref{trho_plot} (c) shows the dependency of the ionization on the laser intensity and ion density. All different cases show differnt evolution of the ionization, but they are all failing to reach the steady state value. 

\subsection{Derivation of the empirical formula}

\begin{figure}[!ht]
    \centering
    \includegraphics[width=.95\linewidth]{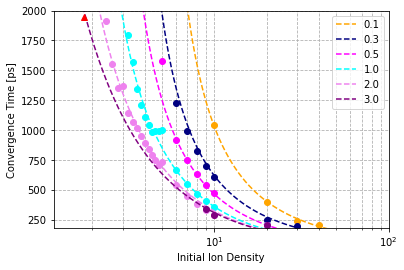}
    \caption{\label{convergencetime-data} Convergence Time as a function of initial ion density in different laser energy. The legend reveals the scaled intensity with the unit of $10^{14}$ $W/cm^{2}$ and the initial ion density is also scaled with $10^{18}$ $cm^{-3}$. This work is marked as the red triangle which is $\sim$ 2000 ps.}
\end{figure}

The convergence time denotes the point at which both the steady-state and time-dependent modes achieve equivalent ionization levels. Within the steady-state mode, equilibrium values are attained for given density and temperature conditions. Under the assumption of constant temperature and density, the time-dependent mode eventually converges to the steady-state mode in every case. This convergence time, reflecting the rate at which this transition occurs, is typically related to the transition rate of the most significant ionization process. The inverse of the rate roughly corresponds to the convergence time. For instance, a collision ionization rate of approximately $10^{12}$ Hz suggests a convergence time of about 1 picosecond.

However, in conventional plasma experimentation, fluctuating temperature and density parameters pose challenges for theoretically estimating convergence. To address this, we validated convergence times for both modes through simulation and introduced a formula for improved comprehension. The Eq.(1) in the main article incorporates parameters such as target density and laser intensity, which are observable in experimental settings.

The data acquisition process entails three primary steps: (1) Conducting TD-L mode simulations for varied laser intensities and initial densities to procure temporal profiles of temperature and density. (2) Performing steady-state simulations for each temporal instance with those density and temperature and compute ionization levels at each instance. (3) Comparing ionization obtained from both steady-state and time-dependent modes to ascertain convergence time. The criterion for convergence is set for the time-dependent value to be within 5\% deviation from the steady-state value. For Ar, this deviation remains below one, which typically does not significantly affect other properties like heat transport.

Results in Fig. \ref{convergencetime-data} demonstrate that higher ion density leads to faster convergence to a steady state due to increased collision frequency, while elevated laser intensity expedites ionization. Fitting ion density and laser intensity data yields a formula akin to Eq.(1) in the main article, with a convergence time of approximately 50 picoseconds, calculated as the root mean square of the dataset.

\end{document}